\documentclass[10pt]{article}
\usepackage{amsmath}
\usepackage{graphicx,psfrag,epsf}
\usepackage{enumerate}
\usepackage{natbib}
\usepackage{url} % not crucial - just used below for the URL 
\usepackage{hyperref}
\usepackage{xurl}
\usepackage{tabularx}
\usepackage{array}
\newcolumntype{C}{>{\centering\arraybackslash}X}
\usepackage{orcidlink}
\usepackage{float}
\usepackage{comment}
\usepackage{booktabs}

\newcommand{\blind}{0}

\begin{document}

\def\spacingset#1{\renewcommand{\baselinestretch}%
{#1}\small\normalsize} \spacingset{1}

%%%%%%%%%%%%%%%%%%%%%%%%%%%%%%%%%%%%%%%%%%%%%%%%%%%%%%%%%%%%%%%%%%%%%%%%%%%%%%

\if0\blind
{
  \title{\bfseries Ownership Networks and Economic Power in the Italian Energy Sector}

  \author{
    Andrea Pannone\thanks{Andrea Pannone and Francesco Giancaterini contributed equally to this work (co-first authors).}
    \\[0.15cm]
    Francesco Giancaterini\footnotemark[1]
    \hspace{0.1cm}\orcidlink{0000-0002-1652-1914}
    \\[0.15cm]
    Tiziano Bacaloni
    \hspace{0.1cm}\orcidlink{0009-0005-9275-7178}
    \\[0.15cm]
    Andrea Bernardini
    \hspace{0.1cm}\orcidlink{0009-0000-5640-8804}
    \\[0.15cm]
    Alessio Abeltino\thanks{Corresponding author: \href{mailto:aabeltino@fub.it}{aabeltino@fub.it}.}
    \hspace{0.1cm}\orcidlink{0000-0002-8966-5425}
    \\[0.4cm]
    \textit{Fondazione Ugo Bordoni, Rome, Italy}
  }

  \date{}

  \maketitle
}
\fi

\if1\blind
{
  \bigskip
  \bigskip
  \bigskip
  \begin{center}
    {\LARGE\bf Strategic Autonomy and Networked Governance
in the Italian Energy Sector}
\end{center}
  \medskip
} \fi

\bigskip
\begin{abstract}
\noindent The energy sector is a cornerstone of national strategic autonomy, yet its increasing financialization has transformed ownership structures into complex networked configurations. This paper investigates the distribution of economic power in the Italian energy sector by introducing two sector-level extensions of the Network Power framework: the Aggregate Network Power Index (A-NPI) and the Aggregate Network Power Flow (A-NPF). Unlike traditional macro-level measures, these indices aggregate firm-level control and influence into a systemic framework that accounts for the relative economic weight of each operator. Applying this framework to the Italian case reveals a “\textit{Governance Paradox}”: while the State retains formal majority ownership, the sector’s deepening reliance on global capital markets and the pervasive presence of common ownership by transnational institutional investors may have altered the relationship between formal public ownership and effective strategic influence. The results reveal a significant concentration of aggregate network power around a relatively limited number of globally diversified institutional investors operating across multiple strategically relevant firms. This configuration raises important questions for European strategic autonomy and infrastructure governance, highlighting potential limitations of traditional Foreign Direct Investment (FDI) screening and antitrust tools in capturing the systemic influence exerted through networked ownership structures.
\end{abstract}

% \text{keywords}
%  \sep Aggregate Network Power Index \sep Aggregate Network Power Flow \sep Energy Sector Financialization \sep Common Ownership \sep Governance Paradox \sep Foreign Direct Investment.
% \end{keywords}

\maketitle

% Main text
\section{Introduction}\label{intro}

Energy systems play a strategic role in modern economies because they directly affect energy security, industrial resilience, and long-term infrastructure coordination. In recent years, the acceleration of the energy transition and the emergence of new geopolitical tensions have further reinforced the strategic relevance of critical energy infrastructures, bringing infrastructure resilience and strategic autonomy back to the center of European energy policy debates (\citealp{Labelle2024}). As a consequence, understanding who controls energy infrastructures and how influence is distributed across the sector has become an increasingly important policy concern.

Historically, the Italian energy sector has been characterized by a strong public presence through direct ownership and regulatory coordination (\citealp{Clo2008}). However, market liberalization, privatization processes, and the growing financialization of large utilities have progressively transformed the governance structure of the sector; see \cite{Florio2010} and \cite{Newell2011}. Major energy firms increasingly operate within hybrid public-private governance structures in which public objectives coexist with market-based financing mechanisms and shareholder-oriented incentives. At the same time, institutional investors and transnational financial actors have acquired a growing role within the shareholder base of critical infrastructures (\citealp{Dudka2023}). As the energy transition requires unprecedented investments in electrification, grid modernization, renewable deployment, and infrastructure resilience, ownership structures play an increasingly important role not only in corporate governance but also in the governance of energy systems and critical infrastructures.

These transformations raise important policy and regulatory challenges. A central issue concerns whether existing tools remain adequate to identify who effectively exercises influence within increasingly complex ownership structures. Traditional policy frameworks generally evaluate concentration and control through direct ownership structures, formal voting rights, or firm-level concentration measures such as the \textit{Herfindahl–Hirschman Index} (HHI) (\citealp{Rhoades1993}). These approaches primarily evaluate concentration at the firm level and may therefore overlook indirect influences emerging from interconnected ownership structures. Similarly, regulatory approaches based on ultimate ownership or Foreign Direct Investment (FDI) screening mechanisms primarily focus on identifiable controlling shareholders and direct acquisitions. However, increasingly interconnected ownership structures may generate forms of indirect influence and systemic concentration that extend beyond formal majority control. In network industries such as energy, influence may emerge through complex ownership configurations distributed across the network rather than through the governance of a single firm alone; see, for example, \cite{Azar2018}.

While a growing body of literature has examined common ownership, corporate governance, and competition in liberalized markets, much less attention has been devoted to how ownership-network structures shape governance outcomes in the energy sector. Existing studies mainly focus on market concentration or firm-level governance, providing limited evidence on how influence is distributed across an entire strategic sector. To our knowledge, no previous study has applied network-based measures of indirect control to evaluate the distribution of influence across the Italian energy sector as a whole.

Against this background, this paper investigates how influence and control are distributed across the Italian energy sector. More specifically, it addresses the following research question: \textit{``to what extent does formal public ownership still correspond to effective control within increasingly financialized and networked energy systems?"}

The paper makes four main contributions. First, it develops a sector-level perspective on energy governance by examining how ownership structures shape the distribution of control and influence across the sector. Second, it examines how control and economic power are distributed across interconnected energy firms operating within a liberalized and capital-intensive system. Third, it introduces two aggregate measures – the Aggregate Network Power Index (A-NPI) and the Aggregate Network Power Flow (A-NPF) – to identify indirect concentration patterns and networked forms of power. Finally, by reconstructing the ownership network linking the main firms operating in the sector, the paper provides a systemic assessment of how economic power propagates through the sector's ownership architecture, discusses the implications of these dynamics for regulatory oversight and infrastructure governance, and illustrates how ownership-network analysis can complement traditional approaches based on direct ownership and market concentration.

The remainder of the paper is organized as follows. Section~\ref{sec:2} reviews the literature on energy governance, financialization, strategic autonomy, and ownership networks. Section~\ref{sec:3} presents the Italian energy sector and discusses its relevance as an empirical setting for the analysis. Section~\ref{sec:4} describes the data, ownership-network reconstruction process, and the Aggregate Network Power framework. Section~\ref{sec:5} presents and discusses the empirical results. Finally, Section~\ref{sec:6} concludes the paper.

\section{Energy Governance and Ownership Structures}\label{sec:2}

Recent debates on energy security and infrastructure resilience have increasingly highlighted the importance of governance structures within liberalized energy systems. At the same time, market liberalization and financialization have progressively transformed the ownership structure of strategic energy infrastructures, increasing the role of institutional investors and transnational financial actors. Building on these developments, this section reviews the literature on energy governance, financialization, ownership structures, and networked influence in strategic sectors.

\subsection{Energy Systems as Strategic Infrastructures}\label{sec:21}

Energy-security research no longer views energy systems simply as supply chains ensuring the availability of fuel and electricity. Instead, energy infrastructures are increasingly understood as critical infrastructures whose stability directly affects industrial production, economic resilience, and national security (\citealp{Cherp2014,Goldthau2014}). This broader perspective reflects the increasing interdependence between energy systems and other critical economic and social activities. Recent geopolitical shocks, particularly the Russia–Ukraine conflict, have further reinforced concerns regarding dependency, infrastructure resilience, and energy security (\citealp{Labelle2024}).

At the same time, the energy transition has increased the importance of infrastructure governance. Electrification, grid modernization, and decarbonization require large-scale investments that must be coordinated over long time horizons (\citealp{Florio2010,Newell2011}). Because the success of these transformations depends on the ability to coordinate investment decisions and manage increasingly complex infrastructure systems, governance arrangements have become a central component of energy-system performance. Recent contributions therefore emphasize that energy security depends not only on the diversification of supply but also on governance capacity, long-term coordination, and the ability to manage systemic vulnerabilities (\citealp{Kuzemko2016,Goldthau2014}). As a result, questions related to ownership, governance, and control have become increasingly central within the study of liberalized energy systems.

These characteristics distinguish energy systems from many other liberalized sectors. Electricity and gas infrastructures operate as highly interconnected network industries requiring continuous system balancing, reliability management, and long-term coordination. Transmission grids and pipeline networks also exhibit strong natural-monopoly characteristics, while infrastructure investments typically involve large fixed costs and long planning horizons. Under these conditions, governance structures may affect not only firm-level performance but also broader dimensions of system resilience, security of supply, and the implementation of long-term decarbonization strategies. For this reason, the distribution of ownership, control, and influence becomes particularly important for understanding how energy systems are governed.

\subsection{Financialization of Energy Infrastructures}\label{sec:22}

Market liberalization transformed the governance structure of European energy systems without eliminating their strategic relevance (\citealp{Florio2010}). Although privatization reduced the direct operational role of the state in several segments of the sector, governments frequently retained ownership stakes in major utilities and transmission infrastructures. Many large energy firms, therefore, evolved into organizations combining public participation with market-based financing structures.
Liberalization also increased the dependence of energy infrastructures on private capital markets and external financing mechanisms. The literature on infrastructure financialization describes this process as the growing integration of infrastructure assets into transnational financial markets and investment portfolios (\citealp{Newell2011}). Utilities and regulated infrastructures became particularly attractive to institutional investors because of their relatively stable revenues and long-term investment profiles. Major utilities increasingly operate under the combined influence of public mandates, shareholder expectations, and market discipline. Under these conditions, effective control may no longer depend exclusively on direct state ownership or formal majority participation. Governance outcomes may instead emerge from the interaction between public actors, institutional investors, and broader ownership configurations surrounding critical infrastructures. 

These tensions become particularly important in the context of the ongoing energy transition. Achieving decarbonization targets requires unprecedented levels of investment in electricity networks, renewable generation capacity, storage systems, and infrastructure modernization. As a result, large energy firms increasingly depend on access to global capital markets to finance long-term investment plans. This dynamic creates what may be described as a capital-dependency loop: the greater the investment requirements associated with the energy transition, the stronger the reliance of major utilities on institutional investors and financial markets. This relationship has important governance implications. As access to external finance becomes increasingly necessary for the implementation of long-term investment plans, institutional investors may acquire a growing role within the governance environment surrounding major energy firms. Consequently, the financing of the energy transition becomes closely intertwined with questions concerning ownership structures, shareholder influence, and corporate governance. While financial integration facilitates access to capital and supports infrastructure expansion, it may also increase the exposure of energy operators to market-based incentives and shareholder expectations. Under these conditions, governance outcomes may be shaped not only by formal ownership structures or public participation, but also by the broader financial environment within which infrastructure operators raise capital and allocate investment resources. The growing dependence on external finance therefore raises broader questions regarding the relationship between financialization, infrastructure governance, and strategic autonomy in liberalized energy systems.

While the financialization literature emphasizes the increasing integration of infrastructure assets into financial markets, a related stream of research focuses more specifically on the role of institutional investors within corporate ownership structures. Energy utilities and network infrastructures have become particularly attractive investment targets because of their stable cash flows, regulated revenues, and substantial financing requirements associated with decarbonization and energy-transition objectives. Recent studies suggest that large institutional investors and asset managers increasingly occupy central positions within corporate ownership structures, allowing them to influence governance outcomes across multiple firms simultaneously (\citealp{Fichtner2017,Aguilera2025}). In infrastructure sectors in particular, these actors may play an important role in mobilizing capital for infrastructure development and modernization while simultaneously exposing firms to financial-market dynamics and shareholder expectations (\citealp{Newell2011}, \citealp{Fichtner2017}, and \citealp{Aguilera2025}). As a consequence, ownership structures increasingly perform a dual function: they constitute mechanisms of corporate governance while also serving as channels through which global financial actors participate in the financing and governance of critical infrastructure systems. Under these conditions, ownership structures become increasingly important not only because they determine who finances infrastructure investments, but also because they shape how influence is distributed across the sector. This evolution further reinforces the importance of understanding how influence propagates through ownership networks in sectors characterized by intensive capital requirements and growing dependence on external finance.

\subsection{Ownership Networks and Indirect Control}\label{sec:23}

Recent research increasingly argues that ownership structures should be analyzed as networked systems extending beyond isolated firm-level relations (see, for example, \citealp{Vitali2011}, \citealp{Brancaccio2018} and \citealp{Mizuno2020}). In complex corporate networks, control does not necessarily coincide with direct majority ownership. Instead, influence may propagate through indirect ownership chains, intermediary financial actors, cross-shareholdings, and coordinated minority participations. As a consequence, formally dispersed ownership structures can still generate highly concentrated patterns of effective control.
This perspective emerged initially within the literature on transnational corporate control and financial networks. Using network-based methodologies, several studies show that a relatively small number of interconnected financial actors occupy structurally central positions within global ownership systems, allowing them to exercise influence that extends far beyond their nominal equity stakes (\citealp{Vitali2011, Garcia2017, Fichtner2017}).
These dynamics appear particularly relevant in liberalized and financialized industries characterized by large institutional investors and fragmented ownership structures. In this context, ownership networks become mechanisms through which influence may propagate across firms and sectors. 

Traditional concentration measures, such as the HHI, concentration ratios, or direct ownership thresholds, often fail to capture these indirect forms of influence because they primarily focus on observable equity shares and market structures rather than on the broader architecture of networked control \cite{Brancaccio2018, Abeltino2026}. As a result, systems that appear competitive at the firm level may still exhibit significant concentration of influence once indirect ownership relations, common ownership structures, and intermediary financial actors are taken into account. Recent network-based approaches attempt to address these limitations by explicitly modeling the propagation of control through ownership networks. Hybrid methodologies combining flow-based and game-theoretic approaches show that influence may depend not only on who ultimately controls a firm, but also on which actors occupy pivotal positions within ownership chains and financial intermediation structures (\citealp{Mizuno2020,Mizuno2023,Pannone2026}).

One important manifestation of these ownership interdependencies emerges through common ownership structures. Recent studies argue that the growing presence of large institutional investors across competing firms may weaken competitive incentives and facilitate forms of strategic alignment even in the absence of explicit coordination (\citealp{Azar2018, Backus2021}). While much of this literature focuses on competition and market outcomes, it also highlights a broader governance issue: financial actors increasingly operate simultaneously across multiple firms, sectors, and infrastructures, creating ownership interdependencies that may weaken the distinction between market competition and coordinated influence. These dynamics may become particularly relevant in liberalized energy markets, where utilities and infrastructure operators often coexist within highly interconnected regulatory and financial environments. This becomes particularly important in strategic sectors, where governance outcomes may emerge from complex interactions between public shareholders, institutional investors, holding companies, and transnational financial intermediaries.

Despite these advances, a clear gap remains in the energy-policy literature. Existing research on common ownership has mainly examined competitive effects, pricing incentives, and market outcomes in conventional product markets. Similarly, studies on corporate ownership networks have primarily focused on global corporate control, financial centralization, or banking and investment structures. By contrast, relatively little attention has been devoted to how networked ownership structures shape the governance of strategic energy infrastructures. In particular, existing approaches have not systematically measured the distribution of effective influence across an entire energy sector composed of multiple interconnected operators, infrastructures, and ownership layers.
This gap is particularly relevant in the energy sector, where governance is not reducible to the control of a single firm. Electricity and gas systems depend on coordinated investment decisions, infrastructure planning, and long-term resilience across multiple operators. Under these conditions, influence may emerge at the sectoral level through common ownership structures, intermediary ownership chains, and the repeated presence of the same investors across multiple firms. As a result, a firm-by-firm analysis may underestimate broader patterns of systemic concentration. This omission is not merely a methodological issue. If influence within strategic energy infrastructures is increasingly exercised through networked ownership structures rather than through direct majority control alone, policymakers may obtain an incomplete picture of governance dynamics when relying exclusively on conventional concentration indicators. As a consequence, existing approaches may underestimate forms of systemic influence that are potentially relevant for infrastructure resilience, strategic autonomy, and long-term energy-transition governance.

This paper addresses this gap by shifting the focus from individual ownership relations to the sectoral distribution of networked economic power. Rather than asking only who controls a specific firm, it asks how influence and control are distributed across the Italian energy sector as an interconnected system of firms and infrastructures. In doing so, the paper contributes to energy-policy research by providing a framework for measuring forms of indirect and systemic influence that remain difficult to detect through standard indicators such as direct ownership shares, HHI-based concentration measures, or conventional FDI-screening tools.

\section{The Italian Energy Sector and Strategic Energy Governance}
\label{sec:3}

The Italian energy sector represents one of the largest and most strategically relevant energy systems in Europe. Italy plays a central role in regional energy coordination through its electricity interconnections, natural-gas infrastructure, and position within Mediterranean energy corridors (\citealp{IEAItaly2023}).

Historically, the sector has been characterized by a strong public presence. Major operators such as \textit{Eni}, \textit{Enel}, \textit{Terna}, and \textit{Snam} originated as state-controlled enterprises and continue to occupy key positions within the national energy system (\citealp{Clo2008}; \citealp{Florio2010}). Although liberalization and privatization reforms have progressively opened the sector to private capital and financial markets, public actors continue to retain important ownership stakes and governance responsibilities in several strategically relevant firms. At the same time, the Italian energy system remains characterized by significant infrastructural interdependencies. Electricity transmission, gas transportation, storage facilities, and distribution networks constitute highly interconnected infrastructure systems that require continuous coordination, long-term investment planning, and reliability management. According to \cite{ARERA2023}, several segments of the gas industry remain relatively concentrated, particularly in infrastructure-related and distribution activities.

These characteristics make the Italian energy sector particularly well-suited to investigating how ownership structures are distributed across strategically relevant infrastructure. The coexistence of public participation, private investors, and interconnected network industries creates an environment in which ownership relations may extend beyond individual firms and become relevant for the governance of the sector as a whole. The following section describes the data sources, ownership-network reconstruction procedure, and Aggregate Network Power framework employed in the empirical analysis.

\section{Data and Ownership-Network Construction} 
\label{sec:4} 
This section describes the data sources, sample construction, and ownership-network reconstruction process employed in the empirical analysis. The objective is to reconstruct the interconnected ownership structure surrounding the main firms operating within the Italian energy sector and to analyze how influence and control propagate across the sectoral governance network. Building upon the institutional and governance dynamics discussed in the previous sections, the analysis adopts an ownership-network perspective that moves beyond isolated firm-level ownership structures and focuses on the broader architecture of interconnected ownership relations linking public entities, institutional investors, municipalities, holding companies, and transnational financial intermediaries. The section is organized as follows. Section~\ref{sec:41} describes the sample selection procedure and the criteria used to identify major firms operating within the Italian energy system. Section~\ref{sec:42} presents the ownership data sources and discusses the treatment of incomplete ownership information. Section~\ref{sec:43} illustrates the ownership-network reconstruction process and the representation of ownership structures as directed weighted networks. Section~\ref{sec:44} reports the main descriptive characteristics of the reconstructed ownership networks. Finally, Section~\ref{sec:45} introduces the Aggregate Network Power framework employed to measure sector-level influence and control across the Italian energy sector.

\subsection{Sample Selection}\label{sec:41}

The empirical analysis focuses on the main firms operating within the Italian energy sector. Since the objective of the paper is to investigate the distribution of influence and control across the Italian energy sector, the sample was constructed to capture the principal operators involved in electricity generation, electricity transmission, gas transportation, gas distribution, storage infrastructures, and integrated utility activities.

The initial identification of firms was based on companies classified under NACE codes 35 (Electricity, gas, steam, and air conditioning supply) and 49.50 (Transport via pipeline). These sectors include the main activities associated with the functioning of the Italian electricity and gas systems and therefore provide a suitable representation of the sector’s infrastructural backbone.
To restrict the analysis to economically significant operators, firms were screened according to their \textit{Earnings Before Interest and Taxes} (EBIT), which is used as a proxy for operational relevance within the sector. The use of EBIT allows comparability across firms characterized by different capital structures and financing conditions, while also excluding very small operators whose role within the broader governance structure of the sector is limited. The sample was subsequently refined by considering the role of firms within the Italian energy system. In this context, strategic relevance refers not only to economic size but also to the role played by firms in the management of critical infrastructures, national energy coordination, and long-term investment activities. This additional criterion is particularly important in the energy sector, where infrastructure relevance does not necessarily coincide with firm size alone. The final sample includes the principal operators active within the Italian energy system, including \textit{Eni}, \textit{Enel}, \textit{Terna}, \textit{Snam}, \textit{Italgas}, \textit{A2A}, \textit{Hera}, \textit{Acea}, \textit{Iren}, \textit{Edison}, \textit{ERG}, \textit{Eviso}, and \textit{Gas Plus}. These firms collectively represent the main infrastructures and organizational configurations characterizing the sector, including state-linked companies, municipal multiutilities, private operators, and firms strongly integrated into international financial markets. The final composition of the sample and the corresponding sectoral classification of the selected firms are reported in Table~\ref{tab:first}.

\begin{table}[ht]
\centering
\renewcommand{\arraystretch}{1.2}

\begin{tabularx}{\textwidth}{
    >{\raggedright\arraybackslash}p{5.5cm}
    >{\centering\arraybackslash}p{2.8cm}
    >{\raggedright\arraybackslash}X
}
\toprule
\textbf{Company Name} & \textbf{NACE} & \textbf{Classification} \\
\midrule
Enel SpA & 35.11 & Production of electricity \\
Eni SpA & 35.22 & Distribution of gaseous fuels through mains \\
Snam SpA & 49.50 & Transport via pipeline \\
Terna Rete Elettrica Nazionale SpA & 35.12 & Transmission of electricity \\
Italgas SpA & 35.22 & Distribution of gaseous fuels through mains \\
A2A SpA & 35.11 & Production of electricity \\
Hera SpA & 35.22 & Distribution of gaseous fuels through mains \\
Acea SpA & 35.23 & Trade of gas through mains \\
Iren SpA & 35.23 & Trade of gas through mains \\
ERG SpA & 35.11 & Production of electricity \\
Eviso SpA & 35.13 & Distribution of electricity \\
\bottomrule
\end{tabularx}

\caption{\textit{Main firms included in the sample and corresponding NACE classification.}}
\label{tab:first}
\end{table}

The selected firms provide a representative picture of the Italian energy sector and of the ownership structures surrounding strategically relevant energy infrastructures. This configuration constitutes an appropriate empirical setting for analyzing how influence and control propagate through interconnected ownership relations within a liberalized and financialized infrastructure system.

\subsection{Ownership Data} 
\label{sec:42} 

Ownership data were obtained from the London Stock Exchange Group (LSEG) Workspace database\footnote{London Stock Exchange Group (LSEG), 2026. LSEG Workspace [database]. Available at: https://www.lseg.com/en/data-analytics/products/workspace [Accessed \today].}. The dataset includes both domestic and foreign shareholders and incorporates ownership relations involving institutional investors, public entities, municipalities, holding companies, financial intermediaries, and private investors. The ownership reconstruction focuses on direct equity participations linking shareholders to firms operating within the Italian energy sector. Particular attention is devoted to the presence of institutional investors and intermediary financial actors operating simultaneously across multiple firms within the sector. As commonly observed in ownership-network studies, raw ownership data remain partially incomplete because disclosure obligations are generally subject to reporting thresholds and transparency requirements. As a consequence, observable ownership shares associated with a given firm do not necessarily sum to 100 percent. Additional limitations emerge from fragmented minority participations, nominee structures, cross-border investment vehicles, and partially undisclosed intermediary ownership arrangements. To address these limitations, the analysis adopts an imputation procedure consistent with the framework proposed by \cite{Pannone2026}. More specifically, missing ownership shares are proportionally redistributed among private investors under a partial coordination assumption. This approach avoids both the unrealistic assumption of perfectly independent minor shareholders and the opposite assumption of fully coordinated ownership structures. In this way, the dataset provides a reconstruction of the broader ownership architecture surrounding the principal firms operating within the Italian energy sector. Rather than focusing exclusively on formal majority shareholders, the analysis captures the interconnected ownership relations linking public actors, institutional investors, municipalities, holding companies, and transnational financial intermediaries across the sector.

\subsection{Ownership-Network Reconstruction}\label{sec:43}

The ownership structures reconstructed in the previous subsection are represented as directed weighted networks, where nodes correspond to firms, shareholders, public entities, institutional investors, municipalities, holding companies, and financial intermediaries. Directed edges represent ownership relations weighted by the corresponding equity participation.

This representation allows the reconstruction of both direct and indirect ownership relations connecting the main actors operating within the Italian energy sector. The resulting ownership network captures the coexistence of public participation, municipal ownership, institutional investors, and transnational financial actors across major energy infrastructures.
Building on this ownership architecture, the following subsection presents the main descriptive characteristics of the reconstructed networks.

\subsection{Descriptive Network Characteristics}\label{sec:44}

\begin{figure}[H]
    \centering
    \includegraphics[width=5.5in]{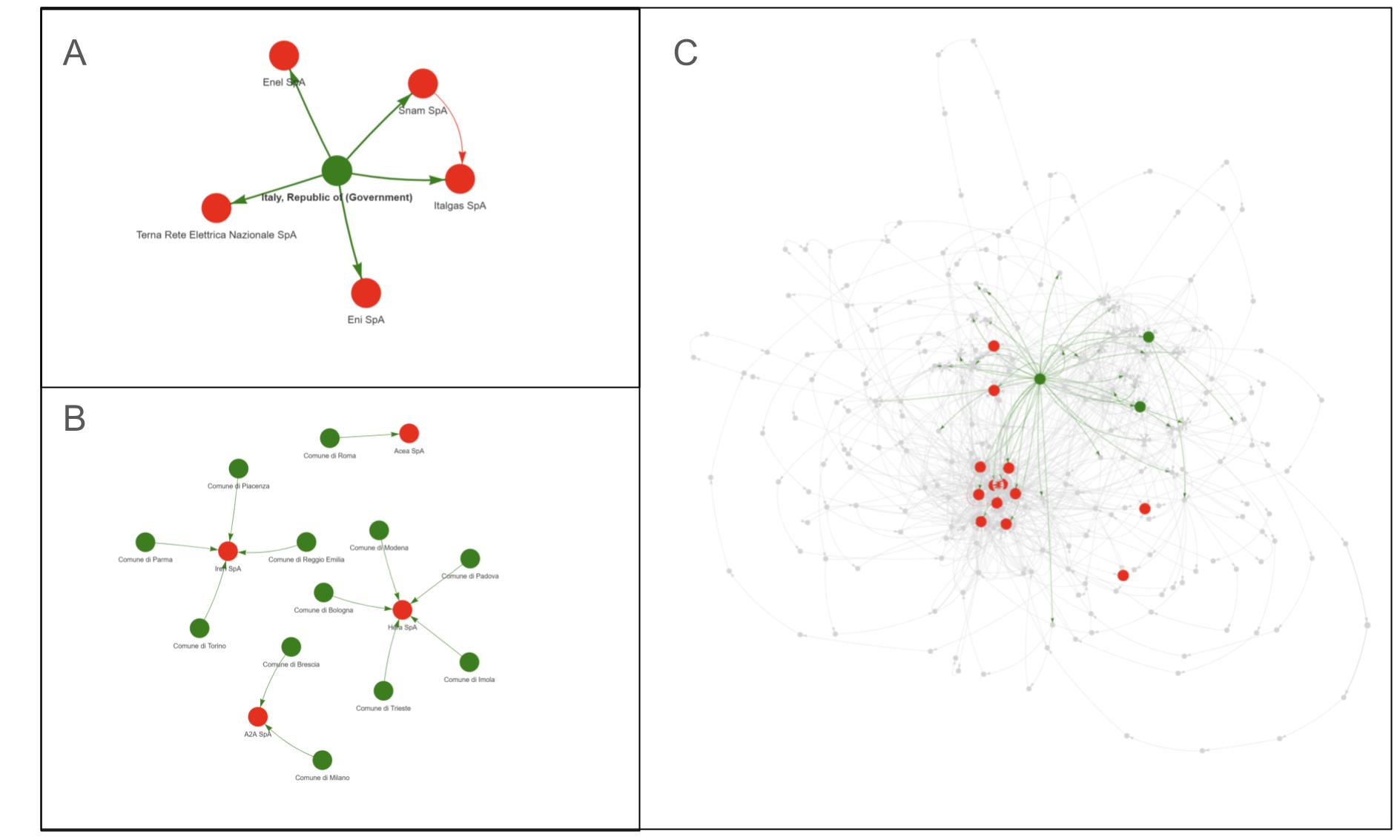}
    \caption{\textit{Shareholding network of major Italian energy firms, based on data retrieved from LSEG as of late 2025. The graphical representation highlights the complex ownership structure of the sector. Panels A and B distinguish public actors, respectively aggregating shareholdings attributable to States (A) and Municipalities (B). Panel C captures the presence of major global asset managers (the “Big Three”). These categories are visualized in green, while the core Italian energy firms are shown in red. Intermediate or non-core corporate entities are represented in grey, illustrating the network of shareholders connected through ownership relations.}\label{fig:first}}
\end{figure}

The ownership-network reconstruction includes public entities, municipalities, institutional investors, holding companies, and transnational financial intermediaries. As shown in Figure~\ref{fig:first}, these actors are connected through a complex system of ownership relations linking major infrastructure operators and financial investors. The network highlights the coexistence of public and private actors within the governance architecture of the Italian energy sector and reveals the recurrent presence of large asset managers across multiple firms, suggesting the existence of overlapping ownership structures that extend beyond individual companies.

\begin{figure}[H]
    \centering
    \includegraphics[width=5.5in]{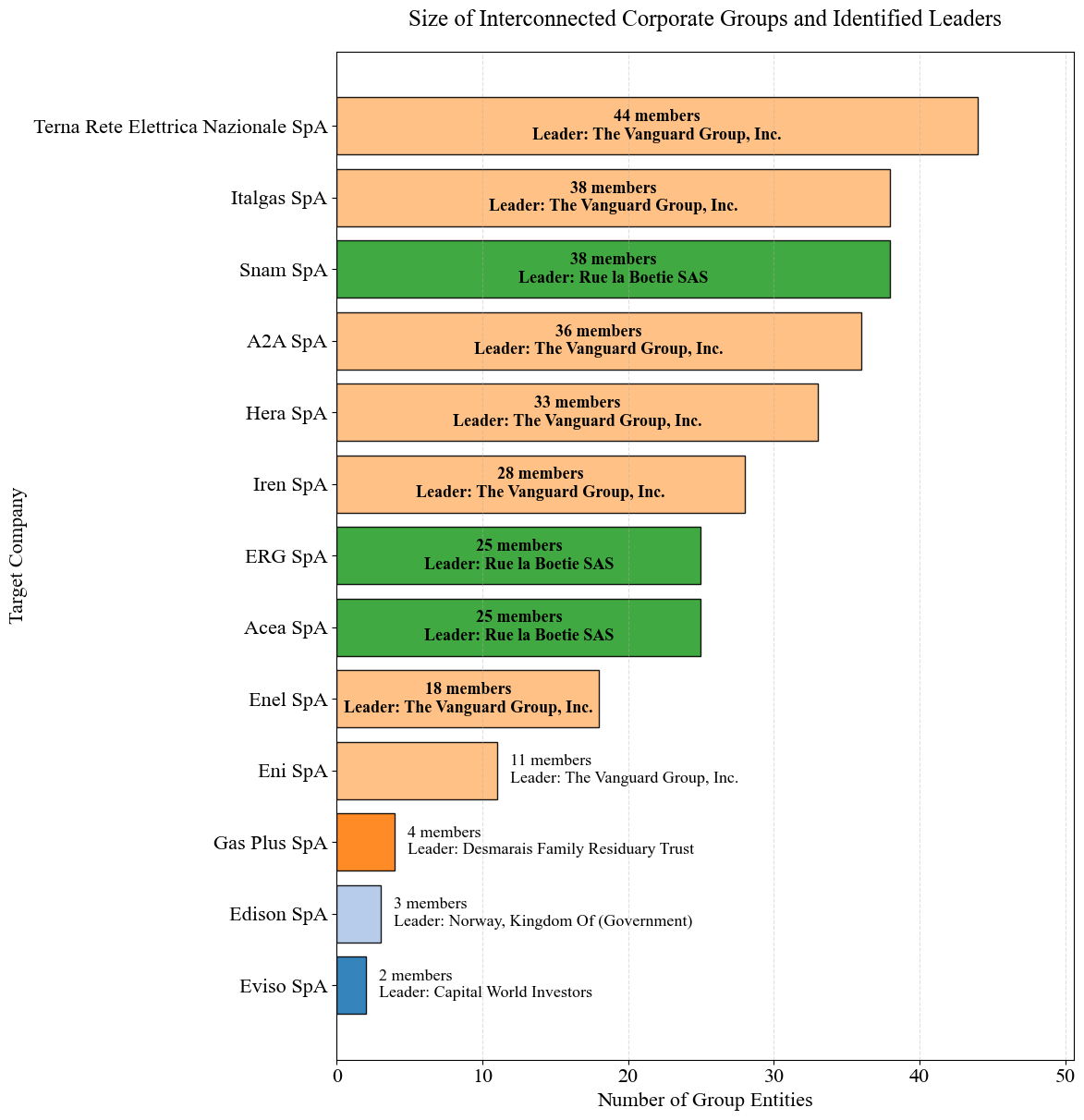}
    \caption{\textit{Number of shareholders belonging to the same corporate group in major energy firms: distribution illustrating the prevalence of interconnected ownership structures. Each colour corresponds to a distinct corporate-group leader, allowing the identification of clustered shareholding configurations within the network.}\label{fig:second}}
\end{figure}

Figure~\ref{fig:second} reports the number of direct shareholders belonging to the same corporate group across the firms included in the sample. The figure provides an overview of the distribution of interconnected ownership structures and highlights the presence of corporate groups represented through multiple entities within the ownership network.

Additional evidence on the structural properties of the reconstructed ownership networks is reported in Table~\ref{tab:second}. Most networks are characterized by a single connected component and relatively low density, reflecting the sparse nature of ownership relations among a large number of shareholders. At the same time, substantial heterogeneity emerges in node connectivity, as reflected by the difference between average and maximum node degrees across the analyzed networks.

Despite their low density, the reconstructed ownership networks exhibit a high degree of overall connectivity. This pattern suggests that ownership relations remain relatively sparse at the local level while still forming an interconnected governance structure through which influence can potentially propagate across multiple actors and ownership layers. The presence of highly connected nodes further indicates that some investors or intermediary entities occupy particularly central positions within the ownership architecture of the sector.

These statistics provide a descriptive overview of the ownership structures that form the basis for the Aggregate Network Power analysis developed in the following subsection.

\begin{table}[ht]
\centering
\small
\setlength{\tabcolsep}{6pt}

\begin{tabularx}{\textwidth}{@{\extracolsep{\fill}}rCCCCCCC@{}}
\toprule
\textbf{Graph} & \textbf{Nodes} & \textbf{Edges} & \textbf{Density} & \textbf{Min Deg} & \textbf{Max Deg} & \textbf{Avg Deg} \\
\midrule
1  & 86  & 98  & 0.0134  & 1 & 24  & 2.28 \\
2  & 281 & 309 & 0.0039  & 1 & 225 & 2.20 \\
3  & 365 & 661 & 0.00498 & 1 & 143 & 3.62 \\
4  & 399 & 755 & 0.00475 & 1 & 135 & 3.78 \\
5  & 391 & 844 & 0.00553 & 1 & 58  & 4.32 \\
6  & 430 & 813 & 0.00441 & 1 & 198 & 3.78 \\
7  & 376 & 701 & 0.00497 & 1 & 105 & 3.73 \\
8  & 200 & 403 & 0.0101  & 1 & 37  & 4.03 \\
9  & 417 & 852 & 0.00491 & 1 & 116 & 4.09 \\
10 & 398 & 844 & 0.00534 & 1 & 109 & 4.24 \\
11 & 421 & 865 & 0.00489 & 1 & 84  & 4.11 \\
12 & 307 & 662 & 0.00705 & 0 & 37  & 4.31 \\
13 & 294 & 648 & 0.00752 & 1 & 37  & 4.41 \\
\bottomrule
\end{tabularx}

\caption{\textit{Descriptive statistics of the ownership networks reconstructed for the firms included in the sample. The table reports the number of nodes and ownership links, network density, connected components, and degree-based measures, providing an overview of the structural characteristics of the ownership architectures underlying the Aggregate Network Power analysis.}}
\label{tab:second}
\end{table}

\subsection{Aggregate Network Power Framework}\label{sec:45}

Traditional measures of market concentration and corporate control, such as the HHI, voting-right concentration, and direct ownership shares, are primarily designed to capture concentration at the firm level. While these indicators provide important information on market structure and ownership distribution, they remain limited in their ability to capture indirect influence, intermediary ownership structures, and the propagation of control through interconnected ownership networks (\citealp{Abeltino2026}). In strategic sectors characterized by complex cross-shareholding structures and overlapping ownership configurations, influence may emerge not only through direct majority participation but also through chains of indirect ownership linking multiple actors across the broader governance structure of the sector.

To address these limitations, recent literature has increasingly adopted network-based approaches to corporate control analysis. In particular, the Network Power framework introduced by \cite{Mizuno2020, Mizuno2023} combines ownership-network analysis with a game-theoretic notion of pivotal control derived from the Shapley-Shubik index (\citealp{Shapley1954}). Within this framework, control is interpreted probabilistically and emerges from the interaction of multiple shareholders whose influence propagates through direct and indirect ownership relations across the network.

The Network Power Index (NPI), introduced by \cite{Mizuno2020}, measures the probability that a given investor ultimately controls a firm once both direct and indirect ownership structures are taken into account. Through Monte Carlo simulations of shareholder coalitions, the approach identifies the investor whose participation becomes pivotal in forming a controlling coalition. The NPI therefore captures the expected distribution of ultimate control across the ownership network.

While the NPI focuses on the identification of ultimate controllers, the Network Power Flow (NPF) introduced by \cite{Mizuno2023} captures how economic influence propagates through the ownership network. Rather than assigning all influence exclusively to the ultimate controller, the NPF distributes influence along the ownership paths linking investors, intermediary entities, and operating firms. Under this interpretation, economic influence may propagate through multiple ownership chains even when an investor does not emerge as the final ultimate owner.

Building upon this framework, \cite{Pannone2026} introduced target-oriented extensions designed to evaluate the governance structure surrounding a specific firm. The Target Network Power Index (T-NPI) measures the probability that an investor ultimately controls a given target firm, while the Target Network Power Flow (T-NPF) captures how influence converges on that firm through the ownership network. These target-based measures are particularly useful in the context of FDI, mergers and acquisitions, and firm-level governance analysis, where the primary objective is to identify the actors exerting influence over a specific company.

However, when the object of analysis is a strategic sector rather than an individual firm, a target-oriented perspective may provide only a partial representation of the effective distribution of influence. In sectors such as energy, sector-wide influence may emerge not from the governance of a single operator, but from the broader ownership configuration linking multiple interconnected firms operating within the same infrastructure system. Under these conditions, systemic influence may arise through the distributed accumulation of minority participations across several strategically relevant firms.

To capture this broader configuration of sector-level influence, this paper introduces two aggregate extensions of the Network Power framework: the A-NPI and the A-NPF. These measures generalize the target-based framework from a single firm to a set of firms belonging to the same economic system.

Let $V$ denote the set of firms included in the analyzed sector. For each investor $i$ and firm $j$, let $T\text{-}NPI_{i \to j}$ denote the probability that investor $i$ ultimately controls firm $j$. The Aggregate Network Power Index is defined as

\begin{equation*}
A\text{-}NPI_i(V) = \sum_{j \in V} T\text{-}NPI_{i \to j}.
\end{equation*}

The A-NPI can therefore be interpreted as the expected number of firms within the analyzed system falling under the ultimate control of investor $i$.

Since firms may differ substantially in economic relevance, the aggregate framework can be extended by introducing firm-level economic weights. Let $v_j \geq 0$ denote an economic size variable associated with firm $j$, such as total assets, revenues, market capitalization, or value added. Normalized weights are defined as

\begin{equation*}
w_j = \frac{v_j}{\sum_{k \in V} v_k},
\qquad \sum_{j \in V} w_j = 1.
\end{equation*}

\noindent Using these weights, the weighted Aggregate Network Power Index is defined as

\begin{equation*}
A\text{-}NPI_i^{w}(V) = \sum_{j \in V} w_j , T\text{-}NPI_{i \to j}.
\end{equation*}

\noindent While the A-NPI focuses on ultimate control, the Aggregate Network Power Flow captures the propagation of economic influence across the ownership network. Let $T\text{-}NPF_{i \to j}$ denote the Target Network Power Flow associated with investor $i$ and firm $j$. The structural Aggregate Network Power Flow is defined as

\begin{equation*} 
A\text{-}NPF_i(V) = \sum_{j \in V} T\text{-}NPF_{i \to j}. 
\end{equation*} 

\noindent Using the same economic weights defined above, the weighted Aggregate Network Power Flow is given by 
\begin{equation*} 
A\text{-}NPF_i^{w}(V) = \sum_{j \in V} w_j \, T\text{-}NPF_{i \to j}. 
\end{equation*} 
\noindent Unlike the A-NPI, which measures the expected concentration of ultimate control, the A-NPF captures the intensity through which influence propagates across ownership paths linking investors, intermediary entities, and major firms. As a consequence, the A-NPI and A-NPF provide complementary information regarding the governance structure of interconnected infrastructure systems. The first captures the expected concentration of ultimate control, while the second highlights the transmission of influence through the broader ownership architecture of the sector.  

From an interpretative perspective, the aggregate measures extend the logic of the target-oriented framework from a single firm to the sector as a whole. A high A-NPI indicates that an investor is expected to exercise ultimate control over a large share of the sector's economically relevant firms. The measure, therefore, captures the concentration of ultimate control at the sectoral level. By contrast, a high A-NPF indicates that an actor plays a prominent role in the transmission of influence across the ownership architecture of the sector. While the A-NPI identifies investors who are more likely to emerge as ultimate controllers, the A-NPF highlights actors whose influence propagates through multiple ownership chains and intermediary entities. Accordingly, the joint analysis of the two indicators distinguishes between different forms of economic power, ranging from concentrated control to structurally embedded influence. Actors characterized by high values of both indicators combine ultimate control and influence transmission at the sectoral level, whereas differences between A-NPI and A-NPF may reveal governance configurations in which influence extends beyond formal control positions through intermediary ownership structures and network connectivity.

Hence, the T-NPI and T-NPF can be interpreted as special cases of the aggregate formulation in which the analyzed system reduces to a single target firm.

\section{Empirical Results}\label{sec:5}

This section presents the main empirical results obtained from the application of the Aggregate Network Power framework to the Italian energy sector. Building upon the reconstructed ownership networks, the analysis evaluates how influence and control are distributed across strategically relevant energy firms once both direct and indirect ownership relations are taken into account.

The A-NPI and the A-NPF identify both the main ultimate controllers and the actors occupying structurally central positions within the transmission of influence across the sector. Beyond the empirical distribution of ownership power, the analysis also discusses the possible governance implications associated with financial intermediation, common ownership structures, and the growing interaction between public strategic interests and transnational financial actors within critical energy infrastructures.

\subsection{Distribution of Aggregate Network Power}\label{sec:51}

Before discussing the broader governance implications, it is useful to examine how aggregate network power is distributed across the main actors operating within the Italian energy sector. The A-NPI and A-NPF provide complementary information regarding ultimate control and influence transmission, allowing the identification of the investors and organizations occupying the most relevant positions within the sectoral ownership network.

\begin{figure}
    \centering
    \includegraphics[width=5.5in]{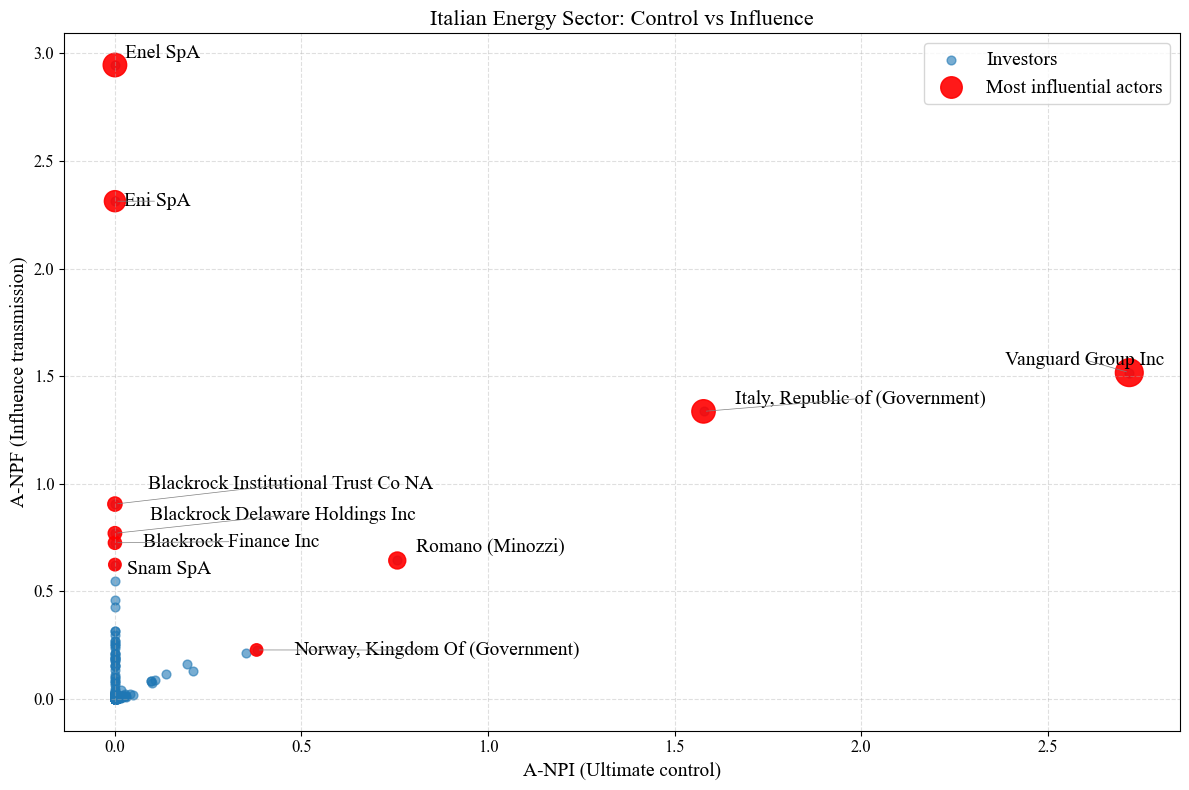}
    \caption{\textit{Distribution of investors in the Italian energy sector according to ultimate control (A-NPI) and influence transmission (A-NPF). Red points denote the most influential actors, identified as those characterized by simultaneously high levels of aggregate control and influence.}\label{fig:third}}
\end{figure}

The main empirical results are summarized in Figure~\ref{fig:third}, which reports the distribution of investors according to their A-NPI and A-NPF. The horizontal axis captures the concentration of ultimate control, while the vertical axis reflects the transmission of influence through ownership chains and intermediary entities. Node size is proportional to the sum of A-NPI and A-NPF. Only actors characterized by significant values of both indicators are displayed for clarity.

Figure~\ref{fig:third} provides a synthetic representation of the distribution of economic power within the ownership network of the Italian energy sector. The joint interpretation of A-NPI and A-NPF highlights different dimensions of power, distinguishing between actors that concentrate ultimate control and those that occupy structurally important positions in the transmission of influence across the network.

The results reveal a pronounced concentration of aggregate network power around a relatively limited number of actors. In particular, \textit{Vanguard Group} emerges as the dominant investor according to both measures. Its position in the upper-right area of Figure~\ref{fig:third} indicates that it combines a high degree of ultimate control with a central role in the propagation of influence throughout the ownership architecture of the sector. The \textit{Italian State} also occupies a prominent position, reflecting its continued presence within several core infrastructures and energy operators.

More importantly, the results suggest that the concentration of sector-level influence is not primarily associated with direct state ownership alone but increasingly with the recurrent presence of globally diversified institutional investors across multiple major operators. Investors such as \textit{Vanguard}, \textit{BlackRock}, and \textit{Norges Bank} simultaneously appear within the ownership structures of electricity producers, gas operators, transmission infrastructures, and integrated utilities. Rather than exercising influence through the control of a single firm, these actors occupy positions that connect multiple infrastructures within the broader ownership network. This configuration indicates that economic power may emerge from the repeated presence of the same financial actors across the sector, generating forms of influence that are not fully captured by conventional firm-level ownership measures. From a governance perspective, this finding suggests that sector-level influence may increasingly derive from the repeated presence of the same institutional investors across multiple strategically relevant operators rather than from direct ownership of individual firms alone. This result does not imply coordinated behavior among investors. Rather, it highlights how networked ownership structures may contribute to the concentration of sector-wide influence even when formal ownership remains dispersed across multiple firms.

The comparison between A-NPI and A-NPF also reveals important differences in the nature of economic power. While investors such as \textit{Vanguard} combine high levels of control and influence, some operating firms display a markedly different profile. Notably, \textit{Enel} and \textit{Eni} are characterized by relatively limited A-NPI values but very high A-NPF scores. This result suggests that their relevance derives less from ultimate control positions and more from their role as central conduits through which influence propagates across the ownership network. More generally, the evidence indicates that influence may extend beyond formal ownership positions through intermediary entities and network connectivity. This distinction would remain largely invisible under conventional ownership indicators based exclusively on direct shareholdings or majority-control thresholds.

To further investigate the nature of this concentration, Table~\ref{tab:third} aggregates network power across broad ownership categories. The results indicate that institutional investors account for the largest share of aggregate network power, representing 65.4$\%$ of total A-NPI and 52.3$\%$ of total A-NPF. Public actors retain a substantial share of aggregate control (28.5$\%$ of A-NPI), reflecting the continued presence of the State within strategically relevant infrastructures. However, their contribution to aggregate influence transmission is considerably lower, accounting for only 8.0$\%$ of total A-NPF. A different pattern emerges for industrial shareholders. Although their contribution to aggregate control is negligible, they account for 37.8$\%$ of total A-NPF. This result suggests that strategically important operating firms occupy structurally central positions within the ownership architecture and play a relevant role in the transmission of influence across the sector. This discrepancy provides the clearest empirical manifestation of the Governance Paradox identified in this paper. Although public actors retain important ownership positions within major firms, their role in the transmission of influence across the broader ownership network is considerably more limited. By contrast, institutional investors account for the majority of aggregate network power despite rarely holding formal majority ownership positions. Their influence derives not primarily from direct control of individual firms, but from their position within the broader ownership network. The evidence, therefore, suggests that formal ownership and effective sector-wide influence do not necessarily coincide within increasingly financialized energy systems, providing empirical support for the Governance Paradox identified in this paper.

\begin{table}
\centering
\small
\setlength{\tabcolsep}{6pt}

\begin{tabularx}{\textwidth}{
    >{\raggedright\arraybackslash}X
    >{\centering\arraybackslash}c
    >{\centering\arraybackslash}c
    >{\centering\arraybackslash}c
    >{\centering\arraybackslash}c
}
\toprule
\textbf{Category} & \textbf{A-NPI} & \textbf{Share A-NPI (\%)} & \textbf{A-NPF} & \textbf{Share A-NPF (\%)} \\
\midrule
Public actors & 2.19 & 28.48 & 1.71 & 8.04 \\
Municipal actors & 0.45 & 5.85 & 0.38 & 1.79 \\
Institutional investors & 5.03 & 65.41 & 11.12 & 52.28 \\
Industrial shareholders & 0.01 & 0.13 & 8.04 & 37.80 \\
Families and foundations & 0.01 & 0.13 & 0.02 & 0.09 \\
\midrule
\textbf{Total} & \textbf{7.69} & \textbf{100.00} & \textbf{21.27} & \textbf{100.00} \\
\bottomrule
\end{tabularx}

\caption{\textit{Distribution of Aggregate Network Power across ownership categories. The table reports both the absolute values and percentage shares of the Aggregate Network Power Index (A-NPI) and Aggregate Network Power Flow (A-NPF).}}
\label{tab:third}
\end{table}

These results indicate that economic power within the Italian energy sector is concentrated around a relatively limited number of actors whose influence extends across multiple strategically relevant firms and infrastructures. More generally, the evidence suggests that economic power depends not only on ownership concentration but also on the position occupied within the broader ownership network. While public ownership remains an important component of the sector's governance architecture, the distribution of aggregate network power indicates that influence is increasingly embedded within wider financial ownership structures. These findings support the view that evaluating strategic control exclusively through direct ownership stakes may provide only a partial representation of governance dynamics within liberalized and financialized energy systems. The broader implications of this ``\textit{Governance Paradox}" are discussed in the following subsection.

\subsection{Public Ownership and Hybrid Governance}\label{sec:52}

The empirical results reveal a ``\textit{Governance Paradox}". On the one hand, the \textit{Italian State} and local public authorities continue to retain important participations in several major firms operating within the sector. On the other hand, the ownership network reconstructed in this paper suggests that the effective governance of the sector is increasingly embedded within broader transnational financial structures characterized by the recurrent presence of large institutional investors across multiple firms. 

The A-NPF is particularly informative in this respect, as it captures how influence propagates through ownership chains and intermediary structures linking investors, holding companies, and operating firms. The empirical evidence suggests that institutional investors occupy structurally central positions within the transmission of influence across the sector, even when they do not necessarily emerge as direct majority owners. This result indicates that sector-wide influence may increasingly derive from the recurrent presence of minority participations distributed across multiple major firms rather than from isolated majority acquisitions alone. These findings are consistent with the literature on common ownership and financialization, which emphasizes how overlapping ownership structures may contribute to the progressive centralization of influence within liberalized markets. Although the observed configurations do not necessarily imply explicit coordination among investors, the recurrent presence of the same institutional actors across major operators may contribute to forms of tacit convergence in infrastructure governance. Under these conditions, the distinction between public and private governance becomes increasingly blurred, as energy infrastructure operators remain formally subject to public oversight while simultaneously embedded within global financial circuits.

\subsection{Financial Intermediation, Strategic Implications and Regulatory Challenges}\label{sec:53}

The observed concentration of aggregate network power also raises broader governance and regulatory questions that extend beyond formal ownership structures. Although these dynamics are not directly captured by the A-NPI and A-NPF, they illustrate how influence may emerge through governance mechanisms that extend beyond formal ownership structures.

Proxy advisory firms such as \textit{Institutional Shareholder Services} (ISS) and \textit{Glass Lewis} increasingly influence shareholder voting outcomes, particularly in firms characterized by fragmented ownership structures and highly diversified institutional investors. Although these actors do not exercise ownership rights directly, their recommendations may contribute to the coordination of voting behavior across multiple firms, introducing an additional layer of governance intermediation within energy infrastructure sectors. A particularly illustrative example emerged during the 2023 board renewal process of \textit{Enel}, Italy’s largest listed utility. Although the \textit{Italian Treasury} remained the company’s largest shareholder with approximately 23.6$\%$ of the capital, the final outcome of the board election was not predetermined. The voting behavior of institutional investors was strongly influenced by proxy-advisor recommendations, illustrating how governance outcomes within strategically relevant infrastructures may increasingly depend on interactions between public ownership structures and globally interconnected financial governance mechanisms (\citealp{Dubois2023}, \citealp{Aguilera2025}, and \citealp{Demirkan2025}).

Similar tensions can be observed in international experiences involving energy infrastructures. In France, the progressive renationalization of \textit{Électricité de France} (EDF) reflected the attempt to regain direct control over long-term energy investments and nuclear infrastructure planning in response to increasing market pressures and financing constraints. Comparable governance tensions have also emerged in the case of \textit{Petrobras}, where the coexistence of public policy objectives and market-oriented financial expectations generated persistent conflicts concerning pricing strategies, investment priorities, and corporate governance structures (\citealp{NemSingh2024, OECD2024}).

These findings suggest that the governance of strategic infrastructures may increasingly depend on networked ownership configurations that are difficult to capture through conventional firm-level concentration measures alone. From a broader political-economy perspective, these dynamics suggest that formal public ownership may increasingly coexist with structurally embedded forms of transnational financial influence. Under these conditions, the governance of critical infrastructures extends beyond traditional distinctions between public and private control, becoming progressively shaped by globally interconnected ownership and financial governance structures operating across liberalized infrastructure systems. Traditional indicators such as the HHI primarily focus on market shares and direct ownership structures, while standard FDI-screening mechanisms generally concentrate on formal acquisitions and majority control. The ownership-network perspective developed in this paper instead highlights how networked influence may propagate through overlapping minority participations, intermediary governance structures, and globally interconnected financial actors operating simultaneously across multiple major firms.

These observations also have implications for regulatory oversight within strategic infrastructure sectors. The ownership-network perspective developed in this paper should not be interpreted as a substitute for existing competition-policy or national-security instruments. Traditional measures such as the HHI, concentration ratios, and foreign direct investment screening mechanisms continue to provide essential information regarding market structure, ownership concentration, and direct acquisitions. However, the results suggest that these approaches may capture only part of the broader governance architecture characterizing increasingly interconnected infrastructure systems. 

Recent international policy discussions point in a similar direction. In particular, \cite{OECD2024} emphasizes that the governance of state-owned and strategically relevant enterprises increasingly takes place within complex ownership environments characterized by hybrid public-private structures, growing financial-market participation, and multiple layers of influence that extend beyond formal ownership rights. From this perspective, understanding who formally owns critical infrastructures may not be sufficient to assess how influence is effectively distributed across strategic sectors. The ownership-network approach developed in this paper contributes to this debate by providing a complementary framework capable of capturing indirect and networked forms of influence that may remain largely invisible under conventional governance indicators.
Ownership-network analysis may therefore represent a complementary analytical tool capable of identifying indirect influence, common ownership structures, intermediary actors, and networked concentration patterns that remain difficult to observe through conventional firm-level indicators alone.

\section{Conclusions and Policy Implications}
\label{sec:6}
The analysis presented in this study highlights a fundamental tension in the ongoing transformation of the Italian energy sector. On the one hand, European policy agendas for the coming years emphasize the importance of strategic autonomy and energy independence as essential tools to reduce exposure to geopolitical shocks (\citealp{Labelle2024} and \citealp{Cherp2014}). On the other hand, the financial structure underpinning the sector points in a different direction, suggesting that the infrastructure supporting the energy transition is increasingly embedded in global financial networks (\citealp{Inderst2010} and \citealp{CarrilCaccia2018}).

The empirical evidence presented in this paper shows that control over key energy companies is not only concentrated but also structurally interconnected across borders. This means that ownership and influence are increasingly shaped by the same set of large international investors operating simultaneously across multiple firms in the sector. As a result, competition between companies may be weakened, as strategic decisions tend to converge around the preferences of a limited number of global financial actors rather than reflecting purely national priorities or competitive market dynamics (\citealp{Azar2018} and \citealp{Backus2021}).

This configuration creates an important policy challenge. While governments may retain formal ownership stakes or regulatory authority, the effective governance of the sector is increasingly influenced by actors whose investment decisions are guided by global portfolio considerations (\citealp{Newell2011} and \citealp{ OECD2024}). In this context, traditional assumptions that equate ownership with control become less reliable, particularly in sectors such as energy, where long-term investment horizons and national security considerations are central (\citealp{OECD2024} and \citealp{Clo2008}).

From a policy perspective, these findings suggest that energy governance can no longer be understood solely in terms of public versus private ownership. Instead, attention must be paid to how investment funds, asset managers, and advisory institutions jointly shape corporate behavior across entire sectors (\citealp{Aguilera2025} and \citealp{Deruytter2022}). The growing interconnection of ownership positions means that influence is exercised through networks rather than through individual firms, creating what can be described as a form of dispersed but coordinated economic power.

Addressing this situation requires a broader reconsideration of existing policy tools. Current frameworks focused on competition policy and foreign investment screening may be insufficient to capture the systemic effects of cross-border ownership networks (\citealp{DOJFTC2023} and \citealp{Rhoades1993}). More sophisticated approaches may be needed to assess how ownership concentration across multiple firms affects market dynamics, investment decisions, and ultimately the direction of infrastructure investment and development (\citealp{Tirole1988} and \citealp{Goldthau2014}).

At a broader level, the findings also raise questions about the governance of critical infrastructure in an era of globalized capital markets (\citealp{Florio2013} and \citealp{NemSingh2024}). Ensuring energy independence may require not only technological innovation and diversification of energy sources, but also a clearer understanding of how financial structures influence strategic decision-making (\citealp{Hirschman1945} and \citealp{ Labelle2024}). Without such an understanding, there is a risk that formal national control coexists with a gradual shift in effective decision-making power toward global financial institutions (\citealp{Kuzemko2016}).

In conclusion, the energy transition should be understood not only as a technological and environmental transformation, but also as a governance challenge (\citealp{Prontera2025} and \citealp{OECD2024}). The increasing centralization of capital and the rise of interconnected ownership networks suggest that new policy instruments will be necessary to ensure that long-term strategic objectives remain aligned with the public interest and territorial energy security (\citealp{Vitali2011}, \citealp{Fichtner2017}, and \citealp{Abeltino2026}).
\section*{Additional information}

\subsection*{Acknowledgments}
\noindent The authors wish to thank Gianni Romano for his technical and methodological assistance in data processing and interpretation, and Professor Takayuki Mizuno for his valuable insights on network-based measures of corporate control. All remaining errors are our own. The authors confirm that they have read and complied with the Taylor \& Francis AI Policy. During the preparation of this manuscript, the authors used ChatGPT (OpenAI) to assist with language editing, including improving grammatical accuracy, correcting spelling errors, and refining phrasing for enhanced clarity and readability. After using this tool, the authors carefully reviewed and revised the manuscript as necessary and take full responsibility for the final content of the publication.

\subsection*{Funding}
\noindent This work was funded under the Agreement signed on December 15, 2023, between the Ministry of Enterprises and Made in Italy and the Fondazione Ugo Bordoni.

% \subsection*{Disclosure statement}
% \noindent Nothing to declare.

\subsection*{Data availability statement}
\noindent Data subject to third-party restrictions. The data that support the findings of this study are available from the London Stock Exchange Group (LSEG). Restrictions apply to the availability of these data, which were used under license for this study.

% Use the style of numbering in square brackets.
% If nothing is used, default style will be taken.
%\begin{enumerate}[a)]
%\item 
%\item 
%\item 
%\end{enumerate}  

%\begin{itemize}
%\item 
%\item 
%\item 
%\end{itemize}  

%\begin{description}
%\item[]
%\item[] 
%\item[] 
%\end{description}  

% Uncomment and use as the case may be
%\begin{theorem} 
%\end{theorem}

% Uncomment and use as the case may be
%\begin{lemma} 
%\end{lemma}

% \appendix

\section*{Appendix A: Algorithm for the Identification of Dominant Corporate Groups}\label{A}
To identify coordinated ownership structures within each company’s shareholder registry, a network-based algorithm was developed using the Python library NetworkX. The ownership structure is modeled as a directed graph, where nodes represent corporate entities and edges represent ownership relations. In this representation, an edge from node A to node B indicates that A holds a share in B. Each edge may also include a weight corresponding to the percentage of equity held.

The objective of the algorithm is to identify the corporate group whose entities collectively control the largest number of direct shareholders of a given target company. This procedure allows the detection of potential coordinated ownership structures in which formally distinct entities belong to the same control chain.

The algorithm operates in the following steps:
\begin{enumerate}
    \item Identification of the target company.
    For each analyzed company, the algorithm retrieves the node corresponding to the company under analysis (the target).
    \item Extraction of direct shareholders.
    All direct shareholders of the target are identified by selecting the predecessor nodes in the graph, i.e., all entities that have an ownership edge pointing to the target.
    \item Identification of potential group leaders.
    The algorithm identifies candidate group leaders as nodes that:
    \begin{itemize}
        \item has no incoming ownership edges (in-degree equal to zero), and
        \item are ancestors of the target company within the ownership network.
    \end{itemize}
    These nodes represent potential ultimate controlling entities at the top of ownership chains.
    \item Reconstruction of corporate groups.
    For each candidate leader, the algorithm computes all of its descendants in the network, which represent the entities belonging to that leader’s ownership chain.
    \item Identification of group-affiliated direct shareholders.
    The algorithm then intersects:
    \begin{itemize}
        \item the set of all entities belonging to the leader’s corporate chain, and
        \item the set of direct shareholders of the target company.
    \end{itemize}
    This step identifies which direct shareholders belong to the same corporate group.
    \item Selection of the dominant group.
    The group leader whose network contains the largest number of direct shareholders of the target company is selected as the dominant corporate group.
    \item Aggregation of ownership information.
    For the selected group, the algorithm returns:
    \begin{itemize}
        \item the group leader,
        \item the list of direct shareholders belonging to the group,
        \item the number of such shareholders, and
        \item the aggregated equity share held by these entities in the target company.
    \end{itemize}
\end{enumerate}

This method allows the identification of clustered ownership patterns in which multiple formally independent entities potentially belong to the same control chain and collectively hold shares in the same company. By focusing on the number of direct shareholders belonging to a common ownership structure, the algorithm captures the density of coordinated participation within corporate networks.






\bibliographystyle{chicago}
\bibliography{Bibliography-MM-MC}
\end{document}